\documentstyle[twocolumn,prl,aps]{revtex}

\begin{document}
\twocolumn[\hsize\textwidth\columnwidth\hsize\csname @twocolumnfalse\endcsname
\draft
\preprint{}
\title{ Destruction of integer quantum Hall effect at strong disorder: A
numerical study}
\author{D.N. Sheng, Z. Y. Weng, and Q. Gao}
\address{Texas Center for Superconductivity and Department of Physics\\
University of Houston, Houston, TX 77204-5506 }
\maketitle
\date{tody}
\begin{abstract}
A breakdown of integer quantum Hall effect (IQHE) at strong disorder is
studied numerically in a lattice model.  We find a generic sequence by which
the  integer  quantum Hall plateaus disappear: higher IQHE
plateaus  always vanish earlier than lower ones. We show that extended
levels between these plateaus do not float up in energy but  keep
merging
together after the disappearance of  plateaus, which  eventually leads to
a localization in the whole system.  We also study this phenomenon  in terms of
topological properties, which provides a simple physical explanation.
\end{abstract}

]

It is an interesting question to ask how the integer quantum Hall effect
vanishes in the strong disorder or weak magnetic field limit. Many years
ago, Khmel'nitzkii\cite{khm} and Laughlin\cite{laughlin} both argued that
extended levels at the
centers of Landau levels would not disappear discontinuously nor merge
together at strong disorder, but rather float up continuously to pass the
Fermi level. This argument is a key scenario
in the global phase diagram proposed by Kivelson, Lee and Zhang\cite{klz} for
 the quantum
Hall effect. Recently a considerable number of experimental
measurements\cite{4,5,6,7,8} have
been devoted to
this issue,  in which  a limited floating-up feature\cite{4,5} is indeed
observed  for lower extended levels. However, such a floating up
seems  not indefinite and experiment\cite{6} suggests that  higher
extended levels eventually  merge with the lowest one at weak magnetic fields.
A direct transition from a higher IQHE state to an insulator state has been
also found.\cite{6,8} These are in contrary to a simple floating up picture and
imply that an overall destruction of IQHE may be more delicate.  Recent
 numerical  study of  finite-size localization length by Liu, Xie,
and Niu\cite{lui} has also suggested no global floating up for
extended levels, although no merge of extended levels
was shown and actually how each of extended levels is
destroyed by disorder remains unclear. Thus the problem of disappearance of the
IQHE needs to be reexamined in a more careful and global way.

In this Letter, we present a numerical study of these issues based on a
 tight-binding model.
A systematic destruction of the integer quantum Hall
plateaus is unambiguously revealed, which exhibits the following sequence:
higher plateaus are found to be
destroyed first by strong disorder in a one-by-one order, and the lowest
plateau
is the last to be gone before the whole system becomes an insulator.
In particular, the lowest extended level located below the last plateau
is not floating up in energy, and
even the critical exponent at such a delocalization point remains essentially
the same as in the full IQHE case. A topological  reason for such a destruction
of the IQHE is given based on the  Chern number which characterizes extended
states\cite{thou2}: nonzero Chern numbers
with opposite sign are moving down from the band center to annihilate those in
the lower-energy extended levels, and eventually a total annihilation of Chern
numbers leads to a global insulating phase at strong disorder.
Furthermore, it is indeed found that a limited floating-up picture for the
lower extended levels shows up when the electron {\it filling number},
instead of the Fermi energy itself,  is fixed with the increase of disorders.
At higher extended levels, we see a direct transition from the IQHE states
to insulator states.

We consider a tight-binding lattice model   of   noninteracting electrons
under a uniform magnetic  field and disorders.  The Hamiltonian is defined as
follows:
\begin{eqnarray*}
H=-\sum_{<ij> } e^{i a_{ij}}c_i^+c_j + H.c. +\sum _i w_i c^+_i c_i . \nonumber
\end{eqnarray*}
Here $c_i^+$ is a  fermionic creation operator, with $<ij>$
referring to two  nearest neighboring sites.  A uniform
magnetic flux per plaquette is
given as  $\phi=\sum _ {\Box} a_{ij}=2\pi/M$, where the  summation runs
over four links around a plaquette. We study the case in which the integer
$M$ is commensurate with the system width L.
And $w_i$ is  a random potential with strength $|w_i|\leq W/2$, and the white
noise limit is considered with no correlations  among different
sites for $w_i$.

Fig. 1 shows an overall picture for the Hall conductance calculated by the Kubo
formula, with a  flux strength $\phi=2\pi/8 $ at a $16\times 16 $ lattice size
(only $E<0$
part is shown in the figure due to the antisymmetry of the Hall conductance on
the
two sides of the band center $E=0$).
In the weak disorder case ($W=1$),  there are three well-defined IQHE plateaus
at $\sigma_H=\nu e^2/h$
($\nu=1,2,3$), corresponding to four Landau levels which are centered at
the jumps of the Hall conductance at $E<0$.  With disorder strength $W$
varying from  $1$ to $6$, we see a systematic destruction of these IQHE
plateaus.  At $W=2$, the third plateau (closest to the band center) begins
to disappear.  An increase of $W$ to
$3$ and $4$ will further result in the disappearance of the second IQHE
plateau, while the lowest  plateau still remains well defined. In the inset of
Fig. 1, such the lowest IQHE plateau at $W=4$ is shown
with sample sizes varying as $8\times 8$, $16\times 16$ and $24\times 24$.
While the quantization of this $\nu=1$ IQHE plateau remains at different
lattice sizes, the region of the transition towards zero is continuously
narrowed with
all the curves crossing at a  fixed-point corresponding to the Hall conductance
$\frac 1 2 e^2/h$, which should be extrapolated to a sharp step of jump in the
thermodynamic limit. This resembles a typical scaling behavior found for the
IQHE\cite{huo1} in the weak disorder case and the lattice-size-independent
fixed-point
represents the mobility edge of the delocalization region. (Later it will be
shown that even the critical exponent at this delocalization energy point
remains essentially the same as in the
full IQHE case.) On the other hand,  the Hall
conductance in the higher energy region at $W=4$ has no more IQHE structure
and is seen to continuously decrease with the increase of the lattice size.  A
breakdown of the lowest  IQHE plateau is eventually found at larger
$W$'s ($W=6$ case is shown in Fig. 1). Here the Hall conductance in the whole
energy region is much less than $e^2/h$, which continuously decreases at
larger sample sizes (from $8\times 8$, $16\times 16$ to $24\times 24$),
indicating
that it will scale
to zero in the thermodynamic limit with all states being localized.

To further inspect the critical behaviors shown in Fig. 1, we use a
different  finite-size scaling method in terms of the Thouless
number.\cite{g-numb}  This method has been
previously used by Ando\cite{ando} to study a similar problem, but
much larger sample sizes (from $16\times 16$ to $96\times 96$) with
more random configurations ($200-30,000$) can be achieved here
by using  Lanczos diagonalization method, in order to accurately decide the
critical points  and exponents.
Fig. 2 shows localization lengths versus energy obtained by a finite-size
scaling calculation.\cite{thou1} Only the data for $W=4$ and $5$ are presented
here where only the lowest IQHE plateau  still remains.  At $W=4$, Fig. 2 shows
that the  localization length follows a scaling behavior $\xi\sim
|E-E_{c1}|^{-x}$ with the exponent $x = 2.4\pm 0.1$ (which is the same as in
the
case when full IQHE plateaus are present at weak disorders\cite{huo}) on the
two sides
of the divergent point  $E_{c1}\simeq -3.40$. Such a
scaling law verifies the existence of a delocalization fixed-point shown in
the inset of Fig. 1 and also confirms that the lowest IQHE plateau at $\nu=1$
is robust as delocalization point $E_{c1}$ is well separated by a
localization region from the higher-energy part. The
localization length $\xi$ exhibits an another dramatic increase which can be
extrapolated as $|E-E_{c2}|^{-4.0}$ when one approaches to $E_{c2}\simeq -2.46$
from below. Such a second delocalization point $E_{c2}$ is more clearly
shown at $W=5$ case in Fig. 2. Above $E_{c2}$ one finds a large but finite
localization length which is rather flat up to the band center and whose
magnitude is reduced at
$W=5$. A recent numerical calculation by using one-parameter finite-size
scaling analysis also suggests such a region to be localized.\cite{xie} Thus
our results show that after losing the IQHE plateaus at
higher energies, original extended levels within this region merge and move
down to $E_{c2}$ at $W=4$ and $5$. Fig. 2 also shows that $E_{c1}$ is
only slightly reduced (from $-3.40$ to $-3.48$) with essentially the same
critical exponent $x$ when $W$ is increased from  $4$ to $5$, whereas $E_{c2}$
is much more quickly moving down (from $-2.46$ to $-2.96$).
Correspondingly the $\nu=1$ IQHE plateau between $E_{c1}$ and $E_{c2}$ becomes
narrower, and eventually
these last two extended levels merge and disappear at a critical $W_c$. We
find that the
localization length becomes finite at $W=6$ in the whole regime, which is
consistent with the conclusion drawn from the Hall conductance calculations,
and places $W_c$ at a range: $5<W_c<6$ . At smaller
disorder strengths, like $W=3$, we also see that
more divergence points  emerge in $\xi$, in correspondence with
more extended levels and additional IQHE  plateaus shown in Fig. 1.

A key thing to understand the above evolution of IQHE plateaus is the
localization-delocalization transition, and it is a well-known fact that
an extended state can be characterized in terms of a topological quantity,
namely, nonzero Chern number\cite{thou2}.
 And the boundary-condition-averaged Hall conductance
is a summation of all the Chern numbers carried by states below
Fermi surface: $<\sigma _H (E_f)>=e^2/h \sum_{\varepsilon_m < E_f} C^{(m)}$.
Here the Chern number $C^{(m)}$ is always an integer. Since its distribution in
the thermodynamic limit decides delocalization regions, one may define
states with nonzero Chern integers as extended states.\cite{huo,she}
Plots of the density of extended states $\rho_{ext}$ are presented in
Fig. 3a with the same flux strength as in Fig. 1. In the weak-disorder case
($W=1$), well-defined peaks of  $\rho_{ext}$ are at centers of Landau-level
bands, separated by localized
regions represented by plateaus in Fig. 1. Widths of these
peaks will approach to zero in the thermodynamic limit.\cite{huo} Total Chern
number for
each of three lower-energy peaks is found to be exactly $+1$, which is the
reason leading to three quantized Hall plateaus at $+1$, $+2$, and $+3$ in unit
of $e^2/h$ shown in Fig. 1, when the Fermi surface is located between these
peaks. The  last peak closest to the band center in Fig. 3 carries a total
Chern number
$-3$, which guarantees that the Hall conductance in Fig. 1
falls back to zero beyond the third plateau when the Fermi energy approaches
$E=0$.
This is a peculiar feature of a lattice model since when the whole band is
half-filled, the Hall conductance has to be zero. With the decrease of the
flux strength and increase of the number of Landau levels or peaks of
$\rho_{ext}$ at the center of the Landau levels, one always sees that the last
peak near the band center carries a large Chern number  with opposite sign  and
a magnitude equal to the total sum of those at lower-energy peaks.

In Fig. 3a, upper two peaks at $W=2$ start to merge due to the
disorder scattering. At $W=3$ they are pretty much mixed together. Because
these two peaks carry total Chern numbers of $+1$ and $-3$, respectively,
an annihilation of these opposite-sign Chern numbers occurs with the
disappearance of the third plateau in Fig. 1, which leads to a substantial
reduction of the magnitude of the Hall conductance in that region. In fact, the
overall
magnitude of the Hall conductance beyond the second plateau is {\it less} than
the value of the second plateau at $W=3$, suggesting that the negative Chern
numbers have already moved down  from the original $-3$ peak and dominated this
region.
Notice that the lowest peak near $-3.4$ in
Fig. 3a still remains separated from the rest spectrum at $W=4$. The inset in
Fig. 3a shows such a peak becomes
narrower and sharper with the increase of lattice sizes. This is consistent
with the finite-size scaling result in Fig. 2 that a well-defined mobility
edge still exists at $E_{c1}=-3.40$. The inset of Fig. 3a also shows a
second peak emerging near $-2.5$, which coincides with $E_{c2}$ in Fig. 2 and
implies that negative Chern numbers would be located in this neighborhood at a
large lattice size, representing a delocalization region beyond which the Hall
conductance jumps back to zero from the $\nu=1$ plateau. Thus when the Fermi
energy is well above $E_{c2}$, the system becomes an insulator with zero Hall
conductance. In other words, a direct transition from original $\nu=2$ IQHE
state to an insulator state is realized here, which is consistent with the
experimental finding.\cite{6,8} Finally, at $W=6$,
the negative Chern numbers reaches to $E_{c1}$ and annihilates the last $+1$
Chern number peak in Fig
3a, where we find a monotonic decrease of $\rho_{ext}$ with larger lattice
sizes which corresponds to a localization in the whole region
as shown by previous analyses.

We do not see any floating up of extended levels before they vanish.  In fact,
extended levels are all moving down at strong disorders
except  $E_{c1}$ which is essentially not changed until this last extended
level is completely annihilated by negative Chern numbers moving down from the
band center. Nevertheless,
a limited floating-up picture may still be seen in the case if the Landau-level
{\it filling number} is fixed as shown in Fig. 3b. If we look
at $\rho_{ext}$ within the lowest Landau level (with the occupation number
$n_L \leq 1$), the peak position does shift towards higher occupation
number at a larger $W$. Since the peak of $\rho_{ext}$ will represent a
mobility edge in the thermodynamic limit, this upward shifting implies that for
a given filling number in the lowest Landau level, the extended states can
float up to pass the Fermi surface when
the disorder is strong enough, leading to an insulating transition. Such a
``floating up'' is due to the reason that
the number of localized states below the lowest extended level at $E_{c1}$ is
increased at stronger disorder, so that when the electron filling number is
fixed the Fermi level is actually moving down, or relatively, the extended
level is shifting up. This effect is indeed consistent with experimental
observations\cite{4,5} for lower extended levels.

Hall conductances at weaker magnetic fields (thus with more Landau levels  as
well as IQHE plateaus) have been also calculated with flux strength
$\phi=2\pi/M$
at $M=11$, $16$ and $24$. All of them exhibit the same generic features as
shown in $M=8$ for the destruction of the IQHE at strong disorder. It suggests
that at large $M$ case (weak-field limit), lower extended levels are
always fixed at
lower energies until higher extended levels move down to merge with them after
higher IQHE plateaus are destroyed in a one-by-one manner. In completing the
present
work, we also became aware of a recent Chern number calculation by Yang and
Bhatt\cite{yang}
at $M=3$, by which they concluded that a ``floating up'' picture is  correct.
Such a floating up of the lowest-delocalization region seen by them is actually
the same as shown in Fig. 3b.  But we point out that
$M=3$ is a very special case where only one plateau ($\nu=1$) exists at $E<0$,
and thus one cannot find  the sequence
of the destruction of higher plateaus as well as the merge of higher
delocalization levels, which are essential in our discussion of  the  global
IQHE evolution at strong disorder.

In conclusion,  our numerical calculations unequivocally demonstrate that the
integer quantum Hall effect in a lattice model is destroyed by strong disorder
by the following sequence: higher integer plateaus closer to band center are
first to vanish when  lower plateaus still remain well defined.  Once an IQHE
plateau disappears,  two delocalization  levels separated by it will merge
together and move down. Since extended states at the band center carry  a total
Chern number with opposite sign as compared to those in the lower
delocalization levels, the sequence of the disappearance of the IQHE plateaus
actually corresponds to that  Chern numbers at the band
center continuously move down towards the band edge  and annihilate those
Chern numbers with opposite sign in all the lower extended levels.  No
floating up in energy  is found for extended levels during such a destruction
procedure. However,  a relative upwards shifting of lower delocalization
levels is indeed  seen if the electron filling number, not the Fermi energy, is
fixed.  Our results thus provide a consistent
explanation of both ``floating up'' as well as direct transitions from higher
IQHE to insulator states  in the experimental measurements.

{\bf Acknowledgments} -The authors would like to thank X. C. Xie for
stimulating discussions and an initial collaboration. The
present work is supported  by TCSUH, and TARP under a grant no.\# 3652182.

Fig. 1.  The Hall conductance $\sigma _H$ as a function of energy E
is plotted for different disorder strength  $W$'s, at a lattice size
$16 \times 16$. The inset shows the evolution of the $\nu=1$ IQHE
plateau at $W=4$ with lattice size varying form $8\times 8$ ($\diamond $),
$16\times 16$ ($\bullet $), to  $24\times 24$ ($+$).

Fig. 2. Localization length  $\xi$ obtained by a finite-size scaling based
on the Thouless number is shown as a function of energy E. Critical behaviors
near divergent points are fitted (dotted curves) by scaling laws (see text).

Fig. 3 (a) The density of extended states $\rho_{ext}$ versus $E$. Disorder
strength $W$'s  are the same as in
Fig. 1  with a lattice size $8 \times 8$ (about 400 random-potential
configurations are used). The inset shows $\rho_{ext}$ at
$W=4$ with lattice sizes varying as $8\times 8$ ($\bullet$), $16\times 16$
($\times$), and $24\times 24$ ($\ast$); (b)  $\rho_{ext}$ versus
Landau-level occupation number $n_L$ with the same parameters
as in (a).


\begin{references}
\bibitem{thou1} Thouless number\cite{g-numb} $g_T(E)=
<\delta E>L^2\rho(E)$,
with $\rho $ as the density of states and $<\delta E>$ represents
the geometric mean value of energy difference of
eigenstates under periodical and antiperiodical
boundary conditions. It satisfies a scaling law:  $g_T=g_0 exp(-L/\xi)$ by
which the localization length $\xi$ can be extrapolated from finite-size
calculations.
\end{references}
\end{document}